# Centimeter-scale superfine three-dimensional printing with femtosecond laser two-photon polymerization

Wei Chu,[1] Yuanxin Tan,[1,2] Peng Wang,[1,3] Jian Xu,[4,6] Wenbo Li,[1,2] Jia Qi,[1,2] Ya Cheng[1,4,5,6]*

[1] State Key Laboratory of High Field Laser Physics, Shanghai Institute of Optics and Fine Mechanics, Chinese Academy of Sciences, Shanghai 201800, China

[2] University of Chinese Academy of Sciences, Beijing 100049, China

[3] School of Physics Science and Engineering, Tongji University, Shanghai 200092, China

[4] State Key Laboratory of Precision Spectroscopy, School of Physics and Materials Science, East China Normal University, Shanghai 200062, China

[5] Collaborative Innovation Center of Extreme Optics, Shanxi University, Taiyuan, Shanxi 030006, China

[6] XXL - The Extreme Optoelectromechanics Laboratory, School of Physics and Materials Science, East China Normal University, Shanghai 200241, China

*ya.cheng@siom.ac.cn

**Abstract**

- Nowadays three-dimensional (3D) printing has been widely used for producing geometrically complex 3D structures from a broad range of materials such as ceramics, metals, polymers, semiconductors, etc. Although it has been demonstrated that a fabrication resolution as high as ~100 nm can be achieved in 3D printing based on two photon polymerization (TPP), the end product size of TPP is typically on millimeter scale limited by the short working distance of high-numerical-aperture focal lens. Here we present a method based on simultaneous spatiotemporal focusing (SSTF) of the femtosecond laser pulses that enables to fabricate centimeter-scale 3D structures of fine features with TPP. We also demonstrate an isotropic spatial resolution which can be continuously tuned in the range of ~10 μm and ~40 μm by only varying the power of femtosecond laser, making this technique extremely flexible and easy to implement. We fabricate several Chinese guardian lions of a maximum height of 0.6 cm and a Terra Cotta Warrior of a height of 1.3 cm using this method.

**Introduction**

Additive manufacturing (AM) has created a revolutionary way of manufacturing by enabling fabrication of geometrically complex there-dimensional (3D) structures of arbitrary shapes and geometries in a single continuous process. As a general principle, AM relies on depositing materials layer after layer to form 3D objects [1]. Although various approaches have been developed to realize high-precision deposition of layers in desired two-dimensional (2D) shapes, AM based on photopolymerization has attracted significant attention for its high spatial resolutions and reasonable fabrication efficiency. In the traditional photopolymerization, liquid photopolymer is cured into the shape of layer with ultraviolet (UV) light [2,3]. The typical layer height achievable with UV polymerization is between ~10 μm and ~100 μm. Meanwhile, the resolution in the horizontal plane of UV polymerization, which depends on the focal spot size of UV beam, can always be different from the layer height, leading to anisotropic spatial resolution.

Alternatively, by scanning a tightly focused femtosecond laser beam in a liquid photopolymer, two-photon polymerization (TPP) can directly produce 3D nanostructures with a spatial resolution far beyond the diffraction limit thanks to a nonlinear threshold effect [4-8]. One of the major challenges in TPP for AM application is its relatively low throughput as TPP is inherently a sequential or series processing technique in which tiny focal spots created by high numerical aperture (NA) objectives must be used to ensure a short Rayleigh length and in turn a high resolution along the axial direction. Even though for some applications which can tolerate moderate sacrifice of fabrication resolution, enlarging the focal spot size in the transverse plane to promote the fabrication throughput is problematic due to a rapid degradation of the resolution in the longitudinal direction. The high NA lenses usually have short working distances, limiting the size of end products fabricated by TPP.

For the reasons given above, it is of critical importance to develop innovative focal schemes which allow to realize isotropic resolutions in TPP independent of the NA of focal lens. Such a focal scheme has been reported by F. He *et. al.* in 2010 by utilizing simultaneous spatiotemporal focusing (SSTF) of the femtosecond laser pulses (SSTF) [9]. Since then, SSTF has been exploited for micromachining and tissue engineering [10-13]. The working mechanism of SSTF has been well documented in Refs. [14,15]. Briefly speaking, implementation of SSTF requires the incident femtosecond pulse to be spatially chirped using a pair of gratings before the pulse enters the focal lens, which dramatically stretches the pulse width and substantially reduce the peak power of the laser beam. After the focal lens, temporal focusing (i.e., shortening of pulse duration) occurs during propagation of the pulse toward the geometric focal point because all the frequency components tend to recombine at the focus. The initial transform-limited pulse of the shortest pulse duration restores itself at the focus. The idea of SSTF focal scheme in femtosecond laser direct writing is borrowed from the SSTF focal scheme developed for widefield bioimaging [16,17]. However, owing to the essential difference in the focal properties required by laser direct writing and widefield imaging, both the focal system and the characteristics of focus are different in these two cases.

In this work, we show that applying SSTF in TPP (i.e., termed as SSTF-TPP hereafter) uniquely allows for producing centimeter-scale 3D structures at spatial resolutions as high as ~10 μm. We demonstrate that tuning of the resolution can be achieved simply by varying the power of femtosecond laser. We investigate the dependence of spatial resolution on the power of femtosecond laser in a quantitative manner. The capacity of SSTF-TPP is confirmed by fabricating complex 3D structures such as Chinese guardian lions and a Terra Cotta Warrior.

**Results and Discussion**

**Control of isotropic resolution** The experimental arrangement of SSTF-TTP is shown in Fig. 1. To demonstrate the 3D isotropic resolution offered by the SSTF-TPP, we first fabricated two arrays of rods, one of which oriented along X direction while the other along Y direction, both embedded in a cube of SU-8 resin as shown in Fig. 2A. The inset of Fig. 2A shows the scanning electron microscopy (SEM) of the fabricated structures. All the rods were written at a fixed scan speed of 400 μm/s, whilst the laser power was varied to write the rods of different cross sectional sizes in the two arrays. Figure 2B-2G shows the optical micrographs of cross sections of the rods fabricated at different femtosecond laser powers. The laser powers below were measured before the grating pair. Considering a total loss of 35% due to the two reflections at the gratings, the laser powers would drop to 65% of the measured powers at the back aperture of the objective lens. In Fig. 2B, the diameters of the rods written with a laser power of 1.5 mW were measured to be ~9.4 μm for both X and Y scan directions, providing a direct evidence on the isotropic fabrication resolution. Then, we gradually raised the laser power to 2 mW, 3 mW and 4 mW in writing each pair of the rods oriented perpendicular to each other, i.e., one in X direction and the other in Y direction. The cross sectional micrographs of the rod pairs fabricated at the increasing laser power are respectively presented in Fig. 2C to 2E. For each pair of the rods, a nearly circular cross sectional shape has been achieved, resulting in various 3D isotropic resolutions of ~13.5 μm, ~20.5 μm and ~27 μm. When the laser power further increased to 5 mW and 6 mW, the cross sections of the fabricated rods slightly deviated from the perfect circular shape, i.e., the rods exhibit diamond-shaped cross sections as shown in Fig. 2F and 2G. At the laser powers of 5 mW and 6 mW, the respective lateral sizes of the rods were measured to be 31.6 μm and 37.7 μm, while the longitudinal sizes were measured to be slightly larger, i.e., 34.5 μm and 44 μm, respectively.

To quantitatively determine the dependence of the fabrication resolution on the power of femtosecond laser, we fabricated a series of rods when we gradually increased the laser power. The lateral and longitudinal cross sectional sizes of the rods are plotted as functions of laser power in Fig. 2H. We observe that in the range of 1.5 mW and 6 mW, the lateral and longitudinal resolutions are well balanced, and both resolutions show nearly linear dependence on the power of femtosecond laser. The mechanism behind the power dependence of the fabrication resolution is the threshold effect in the interaction of femtosecond laser pulses with transparent materials [18,19].

The isotropic 3D resolution was examined in a more straightforward way by writing a series of coils in SU-8 resin at different laser powers, as shown in Fig. 3. In the writing process as illustrated in Fig. 3A, we scanned the SSTF focal spot along a circular trajectory in XY plane while the motion stage was translated along Z direction. Figure 3B to 3D show the optical micrographs of the coils fabricated at various laser power of 3 mW, 4 mW, and 5 mW, respectively. The coils all have a same diameter of 200 μm in XY plane, but they are of different pitches in Z direction. The pitch of the coils in Fig. 3B and 3C is 250 μm, while the pitch of coil in Fig. 3D is 150 μm. The coils show smooth surfaces and circular cross sections by examining at different angles of view. This is consistent with the results in Fig.

2 and provides another unambiguous evidence on the isotropic 3D spatial resolution achieved with SSTF-TTP.

**Fabrication of 3D structures** To showcase the capability of fabricating large-sized 3D structures with SSTF-TPP, we fabricate three Chinese guardian lions of various heights (i.e., 2 mm, 4 mm, and 6 mm) in SU-8 resin. In fabrication of the smallest lion of a height of 2 mm, the structure was sliced into 333 layers with a layer height of 6 μm, and the laser power was fixed at 1.5 mW to obtain the sub-10 μm isotropic resolution in the TPP process. With the scanning speed of 300 μm/s, it took about 5.5 hours to complete the fabrication process. The front-view SEM image of the fabricated lion was shown in Fig. 4A, which faithfully reproduced the model structure as shown in Fig. 4B. To examine the quality of the fabricated structure in greater details, we enlarged the head of the lion to exhibit its details as shown in Fig. 4C. Fig. 4D shows the head part of the model structure from the same angle of view. The peculiar fine features such as the bun structures on the top of the head, the ears, the teeth in the open mouth and the bell on the neck can all been clearly seen in the fabricated lion. The fabricated lion is stable as we do not observe any visible changes in its shape since it was fabricated more than two months ago.

With the tunable resolution, we are able to fabricate larger 3D structures. For this purpose, we kept the layer number unchanged but fabricated the same Chinese guardian lions using higher laser powers. Just by raising the laser power to 3 mW and 4 mW, and resetting the layer heights at 12 μm and 20 μm, respectively, two lions of 4 mm- and 6 mm-heights can be easily fabricated. It took about 7 hrs and 10 hrs to complete the fabrication processes, respectively. The images in Fig. 5(A-C) show the lion of 4 mm-height captured at different angles of view with a digital camera, in which all the fine features are well reproduced. Similar images of the 6 mm-height lion are presented in Fig. 5(D-F), showing no degradation of the fabrication quality despite the increased height of the lion.

At last, a Terra Cotta Warrior with a height of 1.3 cm was fabricated as shown in Fig. 6. The laser power was set at 5 mW and the thickness of slicing layer was set at 30 μm. The entire fabrication process took about 15 hours. The pictures of the Terra Cotta Warrior captured from different angles of view with a digital camera were shown in Fig. 6(B-D). This height is actually limited by the attenuation of the laser pulses in the thick resin and the increasing aberration at greater fabrication depth caused by the difference between the refractive index of air and that of SU-8 resin. We believe that with further effort of aberration correction and sophisticated power compensation, the 3D structures achievable with SSTF-TPP can be made even larger than that have been demonstrated herein.

**Conclusions**

In conclusion, we have applied SSTF-TPP for producing centimeter-scale superfine 3D objectives with isotropic spatial resolutions in the range between ~10 μm and ~40 μm. A large tuning range of spatial resolution has been achieved only by varying the power of femtosecond laser, making this approach easy to implement. This is important for industrial applications in which the troublesome alignment of high-precision optics should be avoided. Another advantage of SSTF-TPP is its relatively high fabrication throughput. In our experiment, fabrication times of several hours are required for producing the centimeter-scale 3D structures because of the low repetition rate (i.e., 1 kHz) of the laser source. Changing to high-repetition-rate laser sources of high average powers will dramatically increase the scan speed, thereby efficiently shortening the fabrication time.

It should be stressed that varying the laser power is not the only choice to tune the resolution of SSTF-TTP. The fabrication resolution can also be tuned by adjusting the width of femtosecond laser beam as we have demonstrated before, which is equivalent to adjusting the NA of the focal system [9-10]. This approach is not chosen in this work as it requires precise realignment of optics, i.e., adjusting the distance between the two gratings, but it should have the advantage of providing larger tuning range of the fabrication resolution. The SSTF-TTP technique breaks the barrier of millimeter-scale end product size in TPP, opening the door for a broad range of superfine 3D printing applications such as micro-electromechanical systems (MEMS), infrared or Terahertz photonics, microfluidics, and 3D bio-printing.

**Materials and Methods**

**Polymer preparation** The polymer used in our experiment is a commercial epoxy-based negative-type resin SU-8. The resin was diluted with acetone with the ratio of 1:1. The resin was then contained in a cuvette and covered with a 170-um-thickness cover glass on the top.

**Pre-baking** Before being solidified with polymerization, SU-8 was pre-baked to evaporate the solvent in the resin. The pre-baking time depends on the size of structures to be fabricated. Typically, for SU-8 resin of 1-cm-thickness, the pre-bake temperature was gradually increased to 95°C at a rate of 5°C/min and then held at this temperature for 30 h.

**Post-baking and Development** After the SSTF-TPP fabrication, the SU-8 resin was post-baked at a temperature of 90°C for 10 min, and then immersed in SU-8 developer for about 6 h to remove the unsolidified liquid resin.

**SSTF-TPP fabrication system** The femtosecond laser amplifier (Libra, Coherent, Inc.) delivered uncompressed 800 nm femtosecond pulses with a bandwidth of ~27 nm and a maximum pulse energy of ~4 mJ at a repetition rate of 1 kHz. The laser beam was attenuated to reduce its pulse energy using a half waveplate combined with a polarizer. In SSTF, the beam width must be reduced which was realized using a telescope beam expander. The laser beam was then directed to a pair of gratings with a groove density of $\sigma$ = 1500 grooves/mm and a blazed angle of 53°. The distance between the two gratings was adjusted to ~730 mm for compensating the temporal dispersion of the uncompressed pulses. The spatially dispersed laser pulses were then focused into polymer material mounted on an XYZ motion stage using a long-working-distance objective lens (Leica, 2 ×, NA =0.35, working distance, 2.0 cm.).

**Data conversion and model realization** The Chinese guardian lion and the Terra Cotta Warrior data were originally saved as stereolithography (STL) files, which were sliced into planes with a constant slice thickness for the subsequent SSTF-TPP fabrication. We scanned the SSTF spot in SU-8 resin along the pre-designed path layer by layer to produce the 3D structures. The scan process was conducted from the bottom of cuvette to the cover glass on the top. Typically, there are two scan strategies in the stereo lithography fabrication: the raster scan and the contour scan. In the former strategy, the whole volume of the structure should be scanned, whilst for the later one, the laser focal spot only scans along the contour profile of the 3D structure. In our experiment, the contour scan had been chosen for its relatively high fabrication efficiency. Meanwhile, we promoted the mechanical strength of the fabricated structure by adding a few reinforcement grid lines within the 3D structures.

**Figures and Tables**

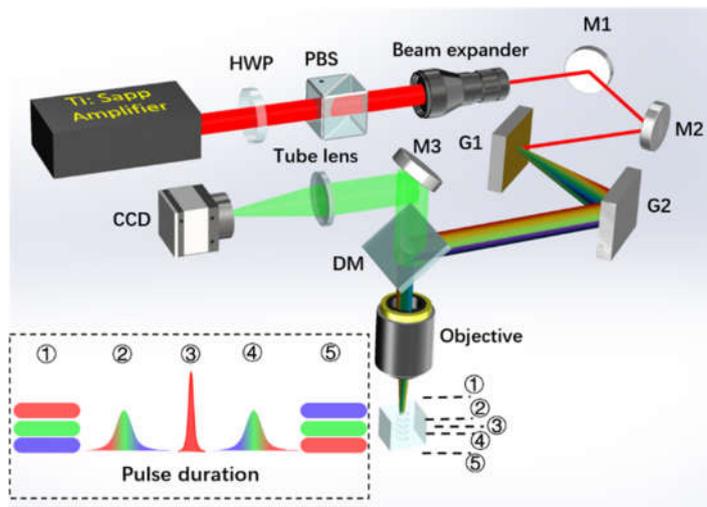

**Fig. 1. Schematic of optical system for SSTF-TPP fabrication.** HWP, half waveplate; PBS, polarizing beam-splitter; M1-M3, reflective mirrors; G1, G2, gratings; DM, dichroic mirror, CCD, charge-coupled device. Inset: The evolution of the pulse duration when the light propagated in focal volume. The spatial position 1, 2 represents the location of before the focal plane; position 3, the focal plane and position 4, 5, after the focal plane. Positions 1 and 5 are far away from the focus, whereas positions 2, and 4 are within the Rayleigh range near the focus.

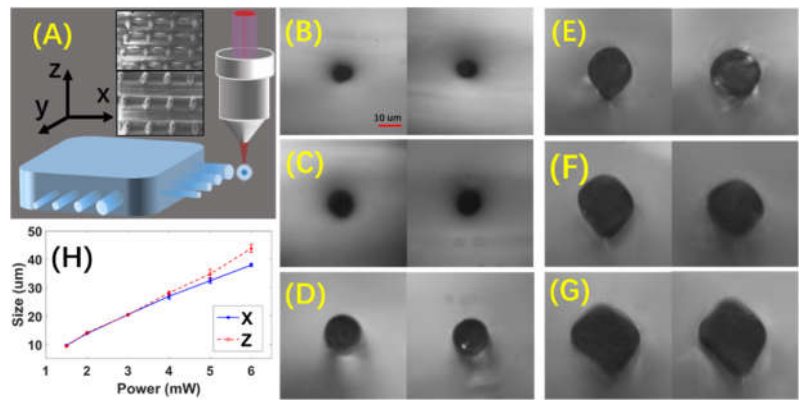

**Fig. 2. Control of 3D isotropic resolution.** (A) Illustration of the designed model structure for measuring the fabrication resolutions at different femtosecond laser powers. Inset: SEM images of the fabricated rods with laser power of 3 mW (upper) and 5 mW (lower). (B-G): the end-sectional-view optical micrographs in XZ (left column) and YZ (right column) with the laser power of 1.5 mW, 2 mW, 3 mW, 4 mW, 5 mW, and 6 mW, respectively. (H): the lateral and longitudinal diameters of the rods plotted as functions of laser power.

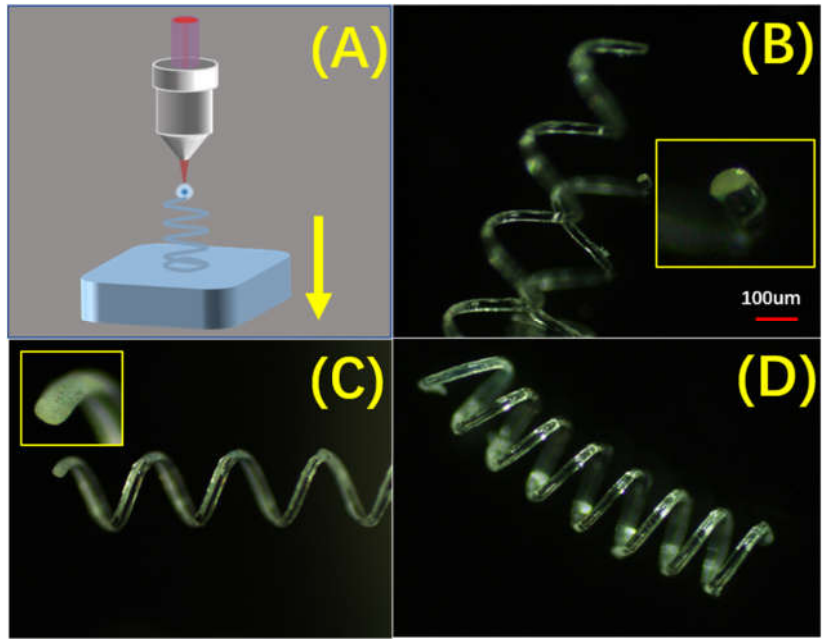

**Fig. 3. Coils writing.** (A) Illustration of the coils fabrication, and the optical micrographs of the fabricated coils produced at different laser powers of (B) 3 mW, (C) 4 mW, and (D) 5 mW. Inset: close-up views of the cross sections at the end of the coils.

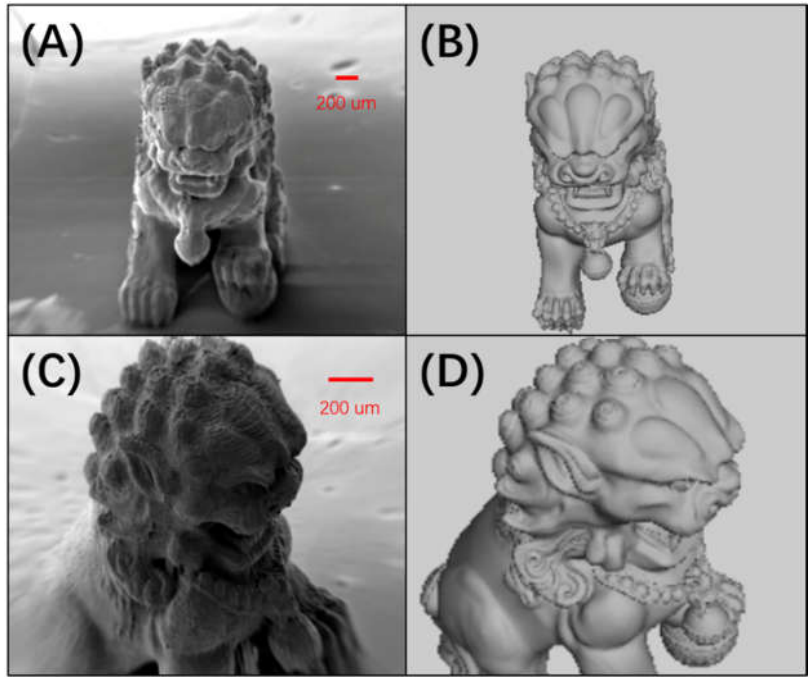

**Fig. 4. The 2-mm-height Chinese lion production.** (A) The SEM image in the front view of the fabricated lion sculpture. (B) The front view of the original model. (C) The SEM image of the head of fabricated lion. (D) The head of the original model exhibited from the same angle of view.

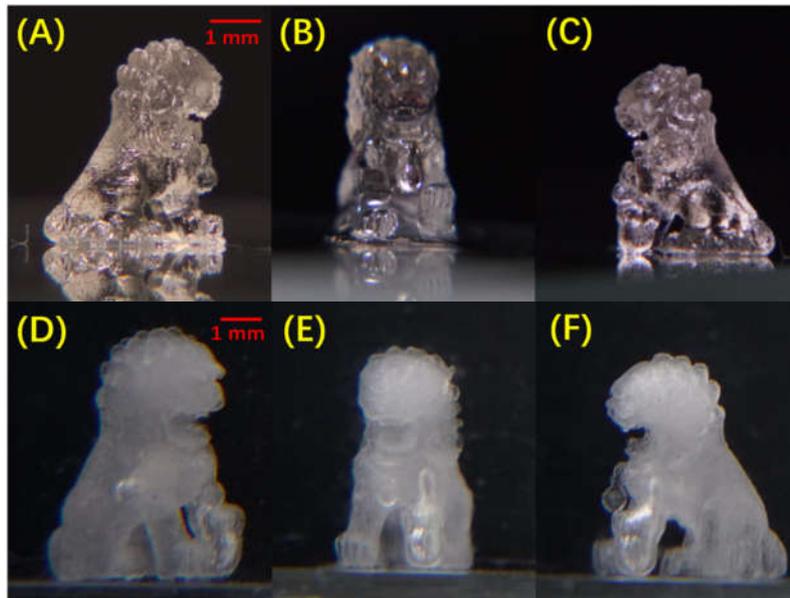

**Fig. 5. The images of Chinese lion sculptures at different angle of views.** (A-C) a lion of 4-mm-height; and (D-F) a lion of 6-mm-height.

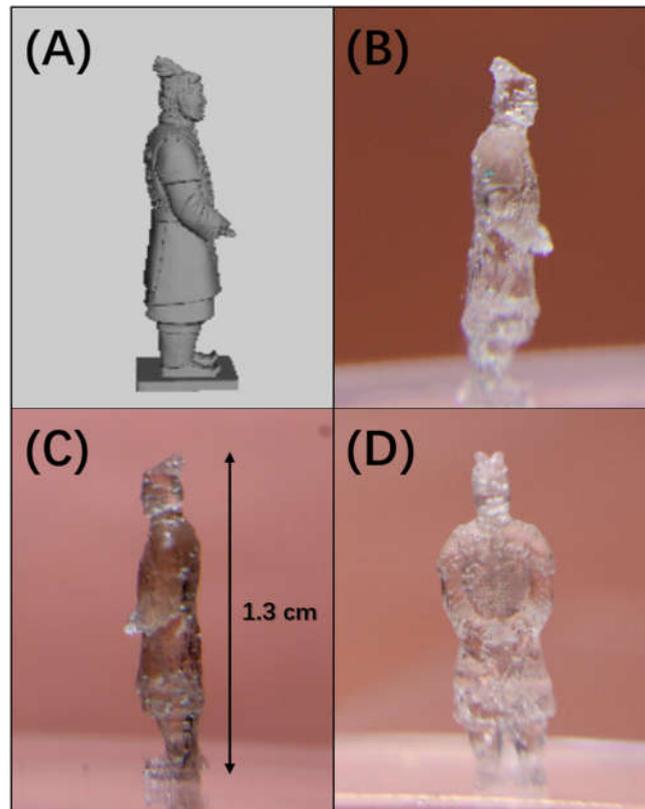

**Fig. 6. Terra Cotta Warrior.** (A) The original model of the Terra Cotta Warrior and the images of the Terra Cotta Warrior sculpture viewed from its right (B), left (C) and front (D) sides as captured by a digital camera.